\documentclass[a4paper,11pt]{article}
\pdfoutput=1 

\usepackage{jcappub} 
\usepackage{tabularx}
\usepackage{booktabs}
\usepackage{multirow}
\usepackage[export]{adjustbox}
\usepackage{floatrow}
\newfloatcommand{capbtabbox}{table}[][3.5in]
\newfloatcommand{capbfigbox}{figure}[][2.3in]
\usepackage{blindtext}
\usepackage{amsmath}
\usepackage{booktabs}
\usepackage[backend=bibtex,style=numeric,natbib=true]{biblatex}
\newcommand{\ie}{i.e.~}

\def\lsim{\mathrel{\raise.3ex\hbox{$<$\kern-.75em\lower1ex\hbox{$\sim$}}}}
\def\gsim{\mathrel{\raise.3ex\hbox{$>$\kern-.75em\lower1ex\hbox{$\sim$}}}}

\begin{document}
\hspace*{110mm}{\large \tt FERMILAB-PUB-22-391-T}

\vskip 0.2in

\title{\boldmath Constraining the Local Pulsar Population with the Cosmic-Ray Positron Fraction}


\author[a,b]{Olivia Meredith Bitter}
\author[a,b,c]{and Dan Hooper}


\affiliation[a]{Fermi National Accelerator Laboratory, Batavia, IL 60510}
\affiliation[b]{University of Chicago, Chicago, IL 60637}
\affiliation[c]{University of Chicago, Kavli Institute for Cosmological Physics, Chicago, IL 60637}

\emailAdd{obitter16@uchicago.com, orcid.org/0000-0002-0600-9095}
\emailAdd{dhooper@fnal.gov, orcid.org/0000-0001-8837-4127}

\abstract{Observations of the TeV halos associated with nearby pulsars indicate that these objects inject significant fluxes of very high-energy electron-positrons pairs into the interstellar medium (ISM), thereby likely providing the dominant contribution to the cosmic-ray positron flux. In this paper, we use the cosmic-ray positron fraction as measured by the AMS-02 Collaboration to constrain the characteristics of the local pulsar population. For reasonable model parameters, we find that we can obtain good agreement with the measured positron fraction up to energies of $E_e \sim 300 \, {\rm GeV}$. At higher energies, the positron fraction is dominated by a small number of pulsars, making it difficult to reliably predict the shape of the expected positron fraction. The low-energy positron spectrum supports the conclusion that pulsars typically transfer approximately $\eta \sim 5-20\%$ of their total spindown power into the production of very high-energy electron-positron pairs, producing a spectrum of such particles with a hard spectral index, $\alpha \sim 1.5-1.7$. Such pulsars typically spindown on a timescale on the order of $\tau \sim 10^4 \, {\rm years}$. Our best fits were obtained for models in which the radio and gamma-ray beams from pulsars are detectable to 28\% and 62\% of surrounding observers, respectively.}

\maketitle


\section{Introduction}
\label{sec:introduction}

Recent observations by the HAWC Collaboration~\cite{HAWC:2021dtl,HAWC:2017kbo,HAWC:2019tcx,HAWC:2020hrt} indicate that young and middle-aged pulsars (and also possibly millisecond pulsars~\cite{Hooper:2021kyp,Hooper:2018fih}) produce significant fluxes of very high-energy gamma rays through the inverse Compton scattering of electrons and positrons~\cite{Hooper_2017,Linden:2017vvb,Profumo:2018fmz,Sudoh:2021avj}. This emission leads to the appearance of ``TeV halos'' surrounding most, if not all, pulsars. Furthermore, the observed characteristics of these TeV halos support the conclusion that these sources are likely responsible for the cosmic-ray positron excess~\cite{Hooper_2017,Hooper_2018} reported by the AMS-02 Collaboration~\cite{Vagelli:2019tqy}, and previously by PAMELA~\cite{PhysRevLett.111.081102}, AMS-01~\cite{Aguilar_2007}, and HEAT~\cite{Barwick_1997} (for earlier work, see Refs.~\cite{Blasi_2009,Profumo:2008ms}).

In previous work, discussions of TeV halos in relation to the cosmic-ray positron fraction tended to focus on the contributions from a small number of nearby and high-spindown power pulsars, such as Geminga and Monogem~\cite{Hooper_2017,Joshi:2017ogv,Fang:2017nww}. It has since become clear, however, that the positrons detected by AMS do not originate from only a few pulsars. Instead, the cosmic-ray positron fraction receives substantial contributions from a significant number of such objects found throughout the surrounding kiloparsecs of the Milky Way~\cite{Cholis:2018izy,Orusa:2021tts}. 

In this paper, we revisit TeV halos as a source of cosmic-ray positrons, using the measured characteristics of the cosmic-ray positron fraction to constrain the properties of the local pulsar population. Toward this goal, we have calculated the cosmic-ray positron flux from the known pulsars contained within the Australia Telescope National Facility (ATNF) pulsar catalog~\cite{Manchester_2005}, taking into account the effects of diffusion, inverse Compton scattering, and synchrotron. We then accounted for those nearby pulsars which have thus far gone undetected, employing a Monte Carlo to simulate this population of sources. Using this Monte Carlo, we have calculated the contribution to the cosmic-ray positron flux from these pulsars, and compared these results to the positron fraction as reported by AMS-02, varying a variety of parameters including the spectral shape of the injected positrons, the timescale for pulsar spindown, the efficiency for electron-positron production, the magnitude of the magnetic field in the interstellar medium (ISM), and the duration of these objects' pulsar wind nebula phase. 

For reasonable choices of these parameters, we have found that it is possible to obtain good agreement with the measured positron fraction, at least up to energies of $E_e \sim 300 \, {\rm GeV}$. This result supports the conclusion that pulsars and their associated TeV halos are likely to be the dominant source of the observed cosmic-ray positron spectrum. At energies above a few hundred GeV, the positron fraction is dominated by a small number of pulsars, making it difficult to reliably predict the shape of the expected positron fraction or to use this high-energy information to draw reliable conclusions pertaining to the Milky Way's broader pulsar population. The low-energy positron spectrum, on the other hand, supports the conclusion that pulsars typically transfer about $\eta \sim 5-20\%$ of their total spindown power into the production of very high-energy electron-positron pairs, and produces a spectrum of such particles with a hard spectral index of $\alpha \sim 1.5-1.7$.

\section{Pulsars as a Source of High-Energy Positrons}
\label{sec:flux}

Once a high-energy electron or positron has escaped from a TeV halo, their transport through the ISM is described by the following diffusion-loss equation:
\begin{align}
\label{eq:diffusionloss}
  \frac{\partial }{\partial t} \frac{dn_e}{dE_e} (E_e,r,t) = \vec{\mathbf{\nabla}} \cdot \bigg[ D(E_e) \vec{\mathbf{\nabla}} \frac{dn_e}{dE_e} (E_e,r,t)\bigg]  +  \hspace{2mm}\frac{\partial }{\partial E_e}  \bigg[\frac{dE_e}{dt}(E_e,r)  \frac{dn_e}{dE_e}(E_e,r,t) \bigg] + \delta(r) Q(E_e,t),
\end{align} 
 where $r$ is the distance from the pulsar, $Q$ characterizes the spectrum and time profile of the injected particles, $dn_e/dE_e$ is the electron and/or positron number density per unit energy, and $E_e$ is the energy of the electrons and positrons. The normalization and energy dependence of the diffusion coefficient, $D$, is constrained by the measurements of various secondary-to-primary ratios in the cosmic-ray spectrum. Throughout this study we will adopt a parameterization given by $D = (2 \times 10^{28} \, {\rm cm}^2/{\rm s}) \, (E_e/{\rm GeV})^{0.4}$~\cite{Vladimirov:2010aq,Hooper:2017tkg}. The energy losses of high-energy electrons and positrons are dominated by a combination of inverse Compton scattering and synchrotron processes, resulting in the following rate:
 \begin{align}
	-\frac{dE_e}{dt} (E_e,r) &= \sum_{i} \frac{4}{3}\sigma_T \rho_i(r) S_i(E_e) \bigg(\frac{E_e}{m_e}\bigg)^2 +  \frac{4}{3}\sigma_T \rho_{\rm mag}(r) \bigg( \frac{E_e}{m_e} \bigg)^2 \\
	&\approx 1.02 \times 10^{-16} \, {\rm GeV/s} \bigg(\frac{E_e}{\rm GeV}\bigg)^2 \times \bigg[\sum_{i} \frac{\rho_i(r)}{ \rm eV/cm^3} S_i(E_e) + 0.224 \,\bigg (\frac{B(r)}{3 \, \mu{\rm  G}}\bigg )^2\bigg ],	\nonumber
	\label{Hooper2}
\end{align}
 where the $\sigma_T$ is the Thomson cross section, $\rho_{\rm mag} \approx 0.224 \, {\rm eV/cm}^3 \times (B_{\rm ISM}/3 \, \mu{\rm G})^2$ is the energy density of the magnetic field, and the sum is performed over different components of the radiation field, consisting of the cosmic microwave background (CMB), infrared emission (IR), starlight (star), and ultraviolet emission (UV). Lastly, the quantity $S_i (E_e)$ is a factor that accounts for the Klein-Nishina suppression of the inverse Compton scattering cross section, and is given by
 \begin{align}
 S_i(E_e) \approx \frac{45 m^2_e/64\pi^2 T^2_i}{(45 m^2_e/64\pi^2 T^2_i)+(E^2_e/m^2_e)}.
 \end{align}
Throughout our analysis, we will follow Ref.~\cite{Hooper_2017} in adopting the following parameters for the radiation model: $\rho_{\rm CMB} = 0.260 \, {\rm eV/cm}^3$, $\rho_{\rm IR} = 0.60 \, {\rm eV/cm}^3$, $\rho_{\rm star} = 0.60 \, {\rm eV/cm}^3$, $\rho_{\rm UV} = 0.10 \, {\rm eV/cm}^3$, $T_{\rm CMB}=2.7 \, {\rm K}$, $T_{\rm IR}=20 \, {\rm K}$, $T_{\rm star}=5000 \, {\rm K}$, and $T_{\rm UV}=20,000 \, {\rm K}$. At very high energies ($E_e \gg m^2_e/T$), inverse Compton scattering is strongly Klein-Nishina suppressed. This is particularly important for the scattering of cosmic-ray electrons and positrons with UV and starlight photons.

For the time profile of the source term in Eq.~\ref{eq:diffusionloss}, we assume that the rate at which high-energy positrons and electrons are injected from a pulsar into the ISM is proportional to the rate at which that pulsar is losing rotational kinetic energy (\ie its spindown power). For the case of magnetic-dipole breaking, a pulsar's spindown power is given by  
\begin{align}
	 \Dot{E}(t) &= -\frac{8 \pi ^4 B^2 R^6}{3 c^3 P(t)^4} \\
    &\approx 1.0  \times  10^{35} \, {\rm erg/s}  \times \bigg(\frac{B}{1.6 \, \times \hspace{1mm} 10^{12} \,{\rm G}}\bigg)^2 \bigg(\frac{R}{15 \,{\rm km}}\bigg)^6 \bigg(\frac{0.23 \,{\rm sec}}{P(t)}\bigg)^4, \nonumber 
\end{align}
where $B$, $R$, and $P$ are the pulsar's magnetic field, radius, and period. Solving this differential equation, we obtain the following expression for the time evolution of the pulsar's period:
\begin{align}
	 P(t) = P_0 \, \bigg(1+\frac{t}{\tau}\bigg)^{\frac{1}{2}},
\end{align}
where $P_0$ is the pulsar's initial period and $\tau$ is the spindown timescale, given by 
\begin{align}
\tau &= \frac{3 c^3 I P_0^2 }{4 \pi^2 B^2 R^6}  \\
    &\approx 9.1 \, \times 10^{3} \hspace{1mm} {\rm yr} \times  \bigg(\frac{1.6 \times 10^{12} \, {\rm G}}{B}\bigg)^2 \bigg( \frac{M}{1.4 \, M_\odot}\bigg)\bigg( \frac{15 \,{\rm km}}{ R}\bigg)^4 \bigg( \frac{P_0}{0.040 \,{\rm sec}}\bigg)^2, \nonumber
\end{align}
where $I$ and $M$ are the moment of inertia and mass of the neutron star.

During a young neutron star's pulsar wind nebulae phase, positrons and electrons can be confined to the region surrounding the object. With this in mind, we set the positron and electron injection rate to zero prior to a time, $t_{\rm PWN}$, which we expect to be on the order of $\mathcal{O}(10^4 \, {\rm yr})$.

To obtain the total source term, $Q$, we multiply the spindown power by a constant efficiency factor, $\eta$, and parameterize the spectral shape using a power-law with an exponential cutoff:
\begin{align}
Q(E_e,t)=
\begin{cases}
0 ,  \,\,\,\,\, \,\,\,\,\,\,\,\,\,\,\,\,\,\,\, \,\,\,\,\, \,\,\,\,\,\,\,\,\,\,\,\,\,\,\,\,\,\,\,\, \,\,\,\,\,\,\,\,\,\,\,\,\,\,\,\,\,\,\,\, \,\,\,\,\,\,\,\,\,\,\,\,\,\,\, t < t_{\rm PWN} \\[10pt]
\eta \, \dot{E}(t) \, Q_0 \, E^{-\alpha}  \, e^{-E_e/E_c}, \,\,\,\,\,\,\,\,\,\, \,\,\,\,\,\,\,\,\,\,\,\,\,\,\, t > t_{\rm PWN} 
\end{cases}
\end{align}
where the normalization coefficient, $Q_0$, is set such that $\eta$ is equal to the fraction of the pulsar's spindown power that goes into the production of positrons and electrons with $E_e > 10 \, {\rm GeV}$ (for all $t > t_{\rm PWN}$):
\begin{align}
\eta = \frac{\int_{10 \, {\rm GeV}} Q(E_e,t) \, E_e \, dE_e}{\dot{E}(t)}.
\end{align}

Putting this together, we solve the diffusion-loss equation as described in Ref.~\cite{Hooper_2017} to find the spectrum of electrons and positrons from a given pulsar at the location of the Solar System. The resulting positron spectrum depends on many parameters, including the distance to the pulsar, the pulsar's age, spindown timescale, initial period, and efficiency, as well as the duration of the pulsar wind nebula phase, the magnitude of the magnetic field in the ISM, and the spectral index and cutoff energy of the injected spectrum. 

Once we have calculated the spectrum of positrons and electrons from the pulsars under consideration, we can combine this information with the spectrum of additional primary electrons (such as those from supernova remnants), as well as the spectra of secondary positrons and electrons that are produced through the propagation of primary cosmic rays. Taken together, this can be used to calculate the cosmic-ray positron fraction, defined as the positron flux divided by the combined flux of electrons and positrons. In calculating the positron fraction, we use the secondary spectra as obtained using the publicly available code GALPROP\footnote{\url{http://galprop.stanford.edu/}.}, and fix the primary spectrum using the measured cosmic-ray electron spectrum, assuming that pulsars produce an identical spectrum of positrons and electrons~\cite{Moskalenko:1997gh,Vladimirov:2010aq}. To account for the finite energy resolution of AMS-02, we convolve the spectrum we calculate in each case with a Gaussian of width $\sigma = 0.02 \, E_e$~\cite{Vagelli:2019tqy}.

\section{Positrons From Known Pulsars}

We begin our analysis by calculating the contribution to the cosmic-ray positron flux from known pulsars; namely those contained within the Australia Telescope National Facility (ATNF) pulsar catalog~\cite{Manchester_2005}. To this end, we consider the 103 ANTF pulsars that are located within 3 kpc of the Solar System and that are younger than $\tau_c < 10^6 \, {\rm years}$.\footnote{The characteristic age of a pulsar is defined by $\tau_c \equiv P/2\dot{P}$. For the case of magnetic dipole braking, this is related to a pulsar's true age, $t$, and spindown timescale according to  $t=\tau_c-\tau$.}

\begin{figure}[t]
\centering \includegraphics[width=120mm]{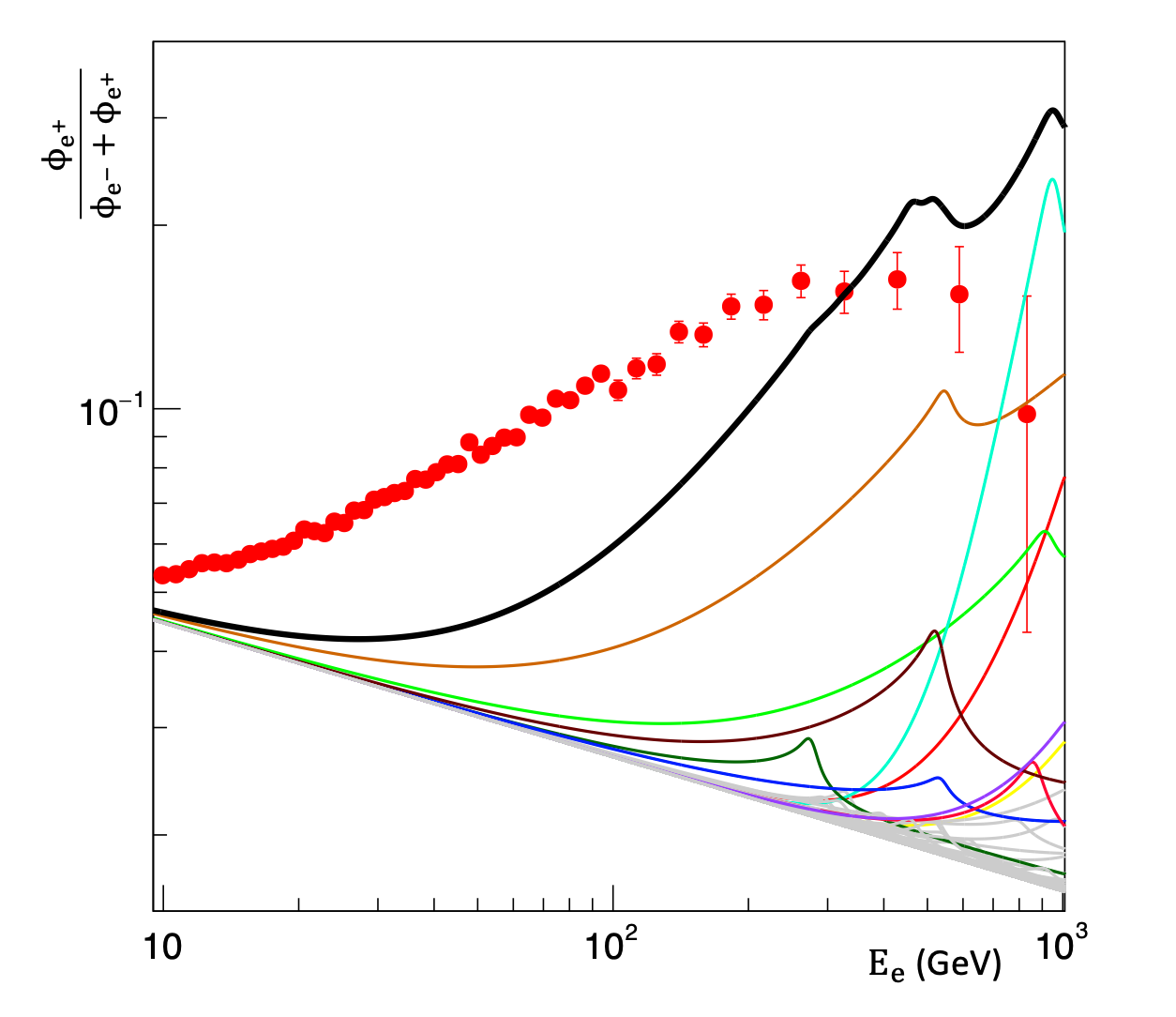}
    \caption{The cosmic-ray positron fraction individually calculated from each of the 103 pulsars in the ATNF catalog that are located within 3 kpc of the Solar System and younger than $\tau_c < 10^6 \, {\rm years}$ (thin colored lines), and from the sum of all 103 of these sources (thick black line). For the color key, see Table~\ref{tab:table1}. In this figure, we have adopted the following parameters: $\alpha=1.5$, $E_c =50 \, {\rm TeV}$, $\eta=0.15$, $\tau=10^4\, {\rm yr}$, $B_{\rm ISM}=3 \, \mu{\rm G}$, and $t_{\rm PWN} = 3 \times 10^4 \, {\rm yr}$. For each individual pulsar, we adopt values of $\tau_c$, $r$, and $P$ as reported in the ATNF catalog.}
    \label{g5}
\end{figure}

In Fig.~\ref{g5}, we plot the contribution to the cosmic-ray positron fraction from each of these 103 pulsars (thin colored lines), and from the sum of all 103 of these sources (thick black line), and compare this to the spectrum measured by the AMS-02 Collaboration~\cite{PhysRevLett.122.101101,PhysRevLett.122.041102,2017ICRC...35.1073D}. Here, we have adopted the following parameters: $\alpha=1.5$, $E_c =50 \, {\rm TeV}$, $\eta=0.15$, $\tau=10^4\, {\rm yr}$, $B_{\rm ISM}=3 \, \mu{\rm G}$, and $t_{\rm PWN} = 3 \times 10^4 \, {\rm yr}$. For each individual pulsar, we adopt values of $\tau_c$, $r$, and $P$ as reported in the ATNF catalog (see Table~\ref{tab:table1}). Keep in mind that this figure only reflects the contributions to the cosmic-ray positron flux from those pulsars contained in the ATNF catalog, which is largely limited to relatively young sources with radio or gamma-ray beams pointed in the direction of the Solar System.

\begin{table}[t]
 \begin{tabularx}{0.66\textwidth}{ccccc}
    \hline\hline \\[-12pt]
 Pulsar Name & $P \, ({\rm s})$ & $d \, ({\rm kpc})$ & $\tau_c \, ({\rm yr})$ & Color\,(Fig.~\ref{g5}) \\
\hline \\[-12pt]

J1302-6350  & 0.048 & 2.63 & $3.22 \times 10^5$& Light Blue/Cyan\\
J1057-5226  & 0.197 &  0.093 & $5.25 \times 10^5$& Orange\\
J0117+5914  & 0.101 & 1.77 & $2.65 \times 10^5$& Red\\
J0633+1746  & 0.237  & 0.190 & $3.32 \times 10^5$& Light Green\\
J2032+4127  & 0.143 & 1.33 & $1.91 \times 10^5$& Violet\\
J0908-4913  & 0.107 & 1.00 & $1.02 \times 10^5$& Yellow\\
J2030+4415  & 0.227 &  0.720 & $5.45 \times 10^5$& Brown\\
J1846+0919  & 0.226 & 1.53 & $3.50 \times 10^5$& Pink\\
J1745-3040  & 0.367 &  0.200 & $5.36 \times 10^5$& Blue\\
J1530-5327  & 0.279 & 1.12 & $9.34 \times 10^5$& Green\\
\hline\hline
\end{tabularx}
\caption{\label{tab:table1} The 10 pulsars contained in the ATNF catalog which contribute the most to the local high-energy cosmic-ray positron flux. For each pulsar, we provide the reported period, distance, and characteristic age. Also given are the colors assigned to each of these pulsars in Fig.~\ref{g5}.}
\end{table}

The locations of the peaks that appear in the spectrum shown in Fig.~\ref{g5} correspond to the energies to which very high-energy positrons can cool in an amount of time equal to the age of the pulsar in question~\cite{Hooper_2017}. For this reason, the older pulsars in this sample tend to exhibit spectral features which peak at relatively low energies. This opens the possibility that individual pulsars could generate distinctive spectral features in the positron spectrum. Conversely, the lack of such features could be used to constrain the characteristics of such pulsars~\cite{Cholis:2017ccs}.

It is interesting to note that the Geminga and Monogem pulsars are responsible for only a modest fraction of the total local positron flux. More specifically, while the contribution from Geminga (PSR J0633+1746, shown in light green in Fig.~\ref{g5}) is among the largest from these pulsars, the positron flux from Monogem does not even make the list of sources presented in Table~\ref{tab:table1}.

\section{Positrons From Unknown Pulsars}

A large fraction of the pulsars in the Milky Way, including many young pulsars located near the Solar System, have not yet been detected, and thus are not contained within the ATNF catalog. To estimate the contribution to the cosmic-ray positron flux from these yet-to-be discovered pulsars, we have employed a Monte Carlo with parameters tuned to reproduce the observed characteristics of the ATNF catalog's pulsar population.

For the locations of the pulsars modelled by our Monte Carlos, we draw from the following spatial distribution~\cite{Lorimer:2003qc}:
\begin{align}
\label{eq:spatial}
	n_{\rm pulsar} \propto R^{2.35} \, e^{-R/1530 \, {\rm pc}} \, e^{-|z|/300 \, \rm pc},
\end{align}
where $R$ and $z$ describe the location of a given pulsar in cylindrical coordinates (we take the distribution to be symmetric with respect to the angular coordinate). We take the solar system to be located at R$=8.25 \, {\rm kpc}$ and $z=0$. For the age of each pulsar, we simply draw from a flat distribution. 

In the calculations performed by our Monte Carlo, we focus on those pulsars which are located within 3 kpc of the Solar System and that are younger than $10^6$ years. While pulsars older than $10^6$ years do contribute to the local positron flux, this portion of the spectrum has a negligible degree of variation from realization-to-realization. Consequently, we use an average output from our Monte Carlo to account for the pulsars in this age range. We also find that pulsars located more than 3 kpc from the Solar System contribute negligibly to the local positron flux, at a level that is smaller than the line width used in our plots.

For each realization of our Monte Carlo, we adopt a value for the pulsar birth rate, $R_{\rm PSR}$, as well as for the spindown timescale, $\tau$. For the spatial distribution described by Eq.~\ref{eq:spatial}, 5.2\% of the pulsar population is located within 3 kpc, so in each realization, we simulate a total of $R_{\rm PSR} \times 0.052 \times 10^6 \simeq 520 \times (R_{\rm PSR}/0.01 \, {\rm yr^{-1}})$ pulsars~\cite{Johnston_2020}. Notice that for reasonable values of $R_{\rm PSR}$, the number of these pulsars is significantly larger than the 103 discussed in the previous section, demonstrating the significant incompleteness of the ATNF catalog.

In order to compare the output of a given Monte Carlo realization to the observed population of young and nearby pulsars, we must establish under what conditions a given pulsar is likely to be detected. In the case of radio detection, the main limiting factor is whether the radio beam is pointed in the direction of the Solar System. To model this, we simply adopt a value for the beaming fraction, $f_{\rm radio}$, which is defined as the probability that a given pulsar produces radio emission in our direction. For this radio emission to be detected, we require both that we reside within the geometry of the radio beam, and that the pulsar's spindown flux is larger than a minimum detectable value, $F_{\rm SD}\equiv \dot{E}/d^2 > F_{\rm radio}^{\rm min}$. In considering the detection of the pulsed gamma-ray emission from pulsars, the corresponding beaming fraction is significantly larger, $f_{\gamma} > f_{\rm radio}$. In this case, however, Fermi is generally only able to detect those pulsars with a relatively high spindown flux, corresponding to $F_{\rm \gamma}^{\rm min} > F_{\rm radio}^{\rm min}$.

\begin{figure}[t]
\centering \includegraphics[width=120mm]{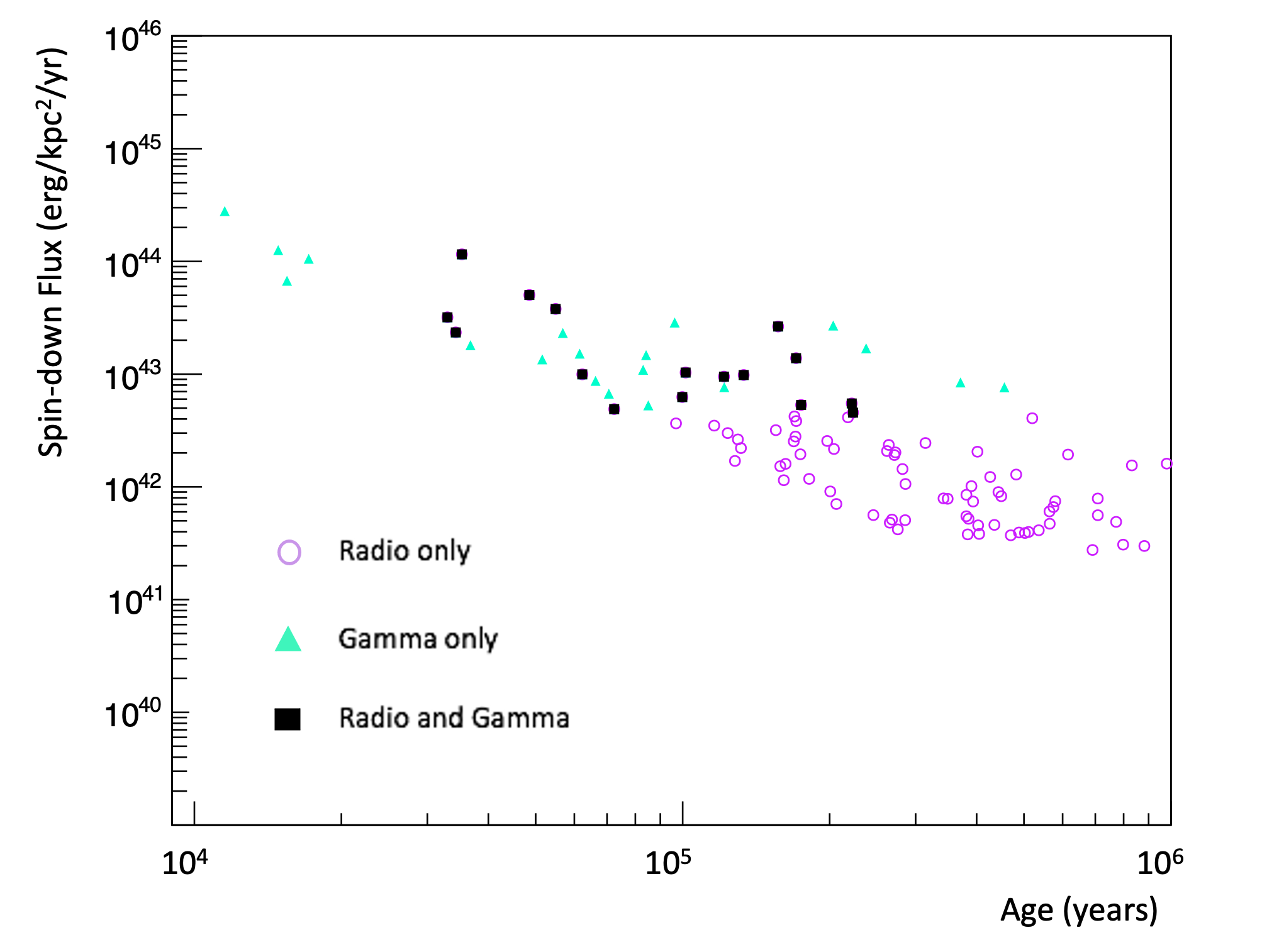}
    \caption{An example of one realization of our Monte Carlo, showing the spin-down flux and age of each detectable pulsar. In this figure, we have adopted the following parameters: $\tau=10^4 \,{\rm yr}$, $P_0=0.042 \, {\rm s}$, $R_{\rm PSR} =0.01 \, {\rm yr}^{-1}$, $f_{\rm radio}=0.28$, $f_{\gamma}=0.62$, $F_{\gamma}^{\rm min} = 4.22 \times 10^{42} \, {\rm erg/kpc}^2/{\rm yr}$, and $F_{\rm radio}^{\rm min} = 2.66 \times 10^{41} \, {\rm erg/kpc}^2/{\rm yr}$. The symbols used reflect whether a given pulsar is detectable by their radio or gamma-ray emission.}
    \label{g9}
\end{figure}

In Fig.~\ref{g9}, we show an example of one realization of our Monte Carlo, plotting the spin-down flux and age of each detectable pulsar. In this figure, we have adopted the following parameters: $\tau=10^4 \,{\rm yr}$, $P_0=0.042 \, {\rm s}$, $R_{\rm PSR} =0.01 \, {\rm yr}^{-1}$, $f_{\rm radio}=0.28$, $f_{\gamma}=0.62$, $F_{\gamma}^{\rm min}$ = 4.22 $\times 10^{42} \, {\rm erg/{kpc}^2/{\rm yr}}$, and $F_{\rm radio}^{\rm min}$ = 2.66  $\times 10^{41} \, {\rm erg/{kpc}^2/{\rm yr}}$. The symbols used indicate whether a given pulsar is deemed to be detectable by their radio or gamma-ray emission.

\begin{figure}[H]
\centering \includegraphics[width=76mm]{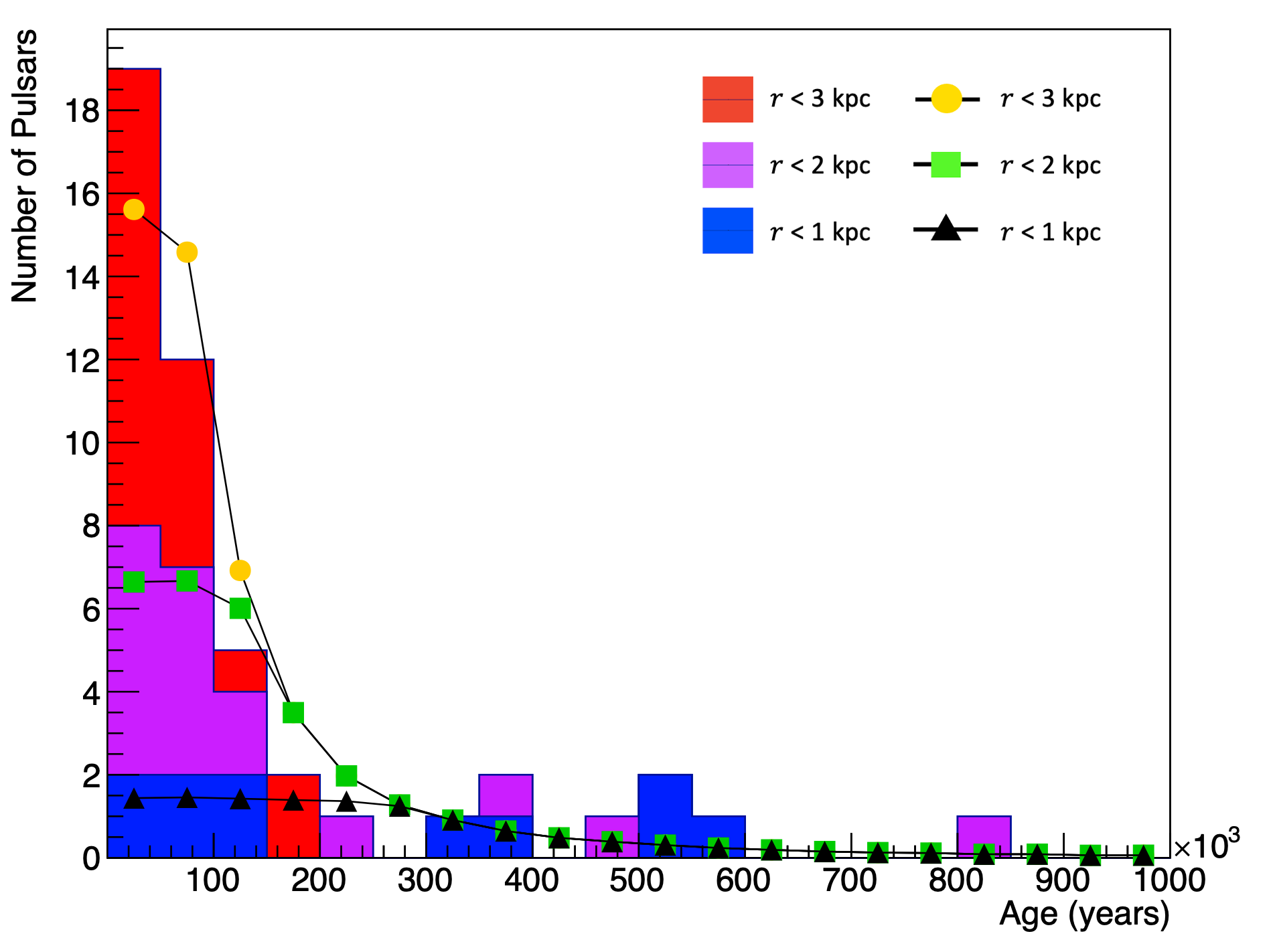} 
\centering \includegraphics[width=76mm]{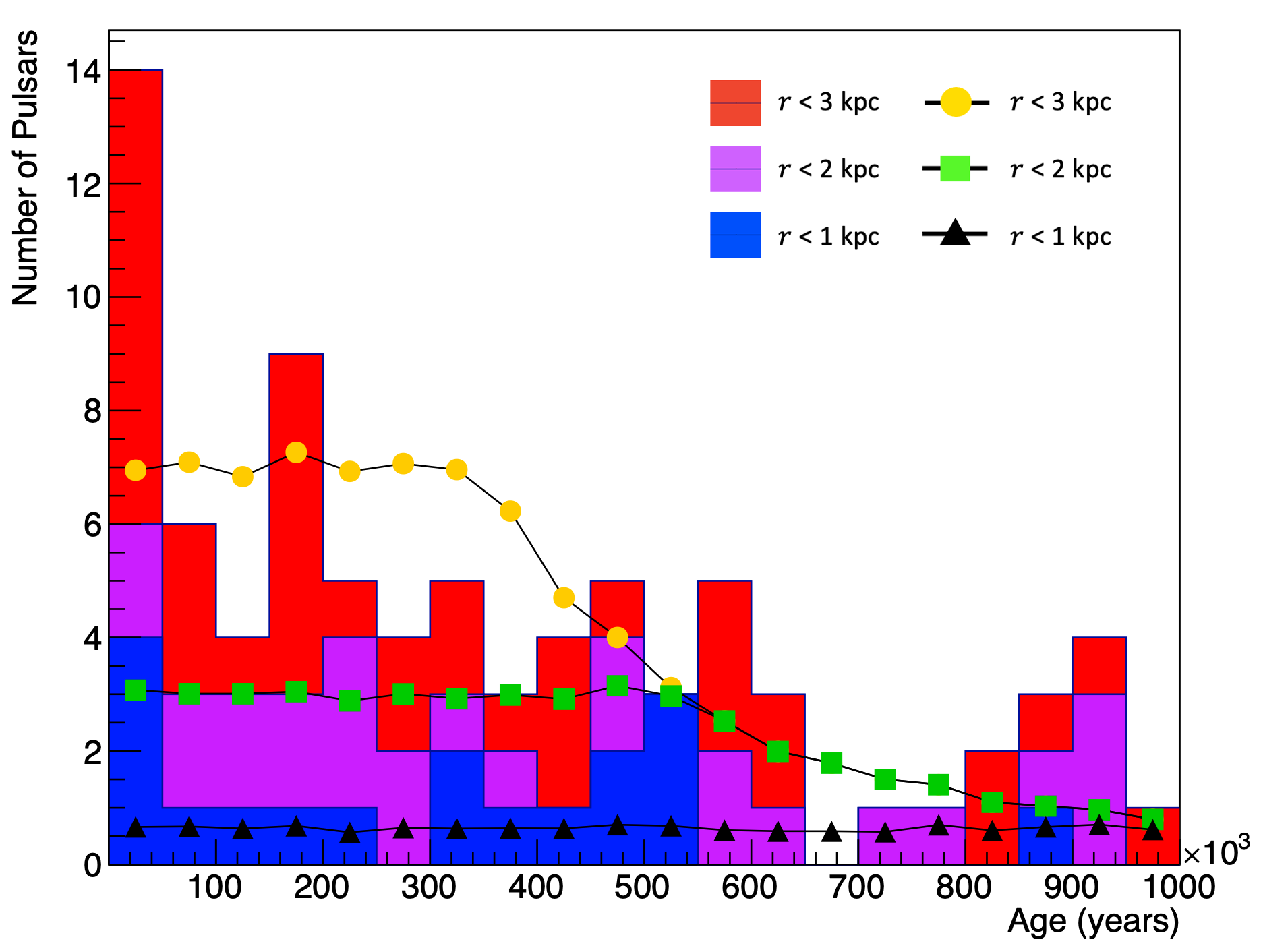} 
    \caption{The distribution of ATNF pulsars detected in gamma-rays (left) and in radio emission (right), according to their age and distance from Earth. The curves represent the distributions predicted by our Monte Carlo, averaged over 1000 realizations, for the case of $\tau=10^4 \, {\rm yr}$, $R_{\rm PSR} = 0.01 \, {\rm yr}^{-1}$, $F_{\gamma}^{\rm min} = 4.22 \times 10^{42}$ ergs/kpc$^2$/yr, $F_{\rm radio}^{\rm min} = 2.66 \times 10^{42}$ ergs/kpc$^2$/yr, $f_{\gamma}=0.62$, and $f_{
\rm radio} = 0.28$.}
    \label{g12}
\end{figure}

To constrain the values of $f_{\gamma}$, $f_{\rm radio}$, $F_{\gamma}^{\rm min}$, and $F_{\rm radio}^{\rm min}$, we compare the observed distribution of detected pulsar ages to the predictions of our Monte Carlo for various combinations of parameter values. These distributions are shown in Fig.~\ref{g12} for the case of $\tau=10^4 \, {\rm yr}$ and $R_{\rm PSR} = 0.01 \, {\rm yr}^{-1}$, using the best-fit parameter values of $F_{\gamma}^{\rm min} = 4.22 \times 10^{42}$ ergs/kpc$^2$/yr, $F_{\rm radio}^{\rm min} = 2.66 \times 10^{42}$ ergs/kpc$^2$/yr, $f_{\gamma}=0.62$, and $f_{\rm radio} = 0.28$. The histograms in these figures represent the number of pulsars detected by Fermi (left frame) and by radio telescopes (right frame), which are then compared to the results of our Monte Carlo, averaged over 1000 realizations. For each choice of parameters, we calculate the Poisson likelihood of obtaining a distribution of pulsars (in terms of age, and in the three distance increments shown) that is consistent with the measured histograms. After repeating this exercise for various parameter values, we identify the best-fit values of $F_{\gamma}^{\rm min}$, $F_{\rm radio}^{\rm min}$, $f_{\gamma}$, and $f_{\rm radio}$ for each value of $\tau$ and $R_{\rm PRB}$ as given in Table~\ref{tab:tableS}. Throughout the remainder of our analysis, we will adopt the best-fit values of $F_{\gamma}^{\rm min}$, $F_{\rm radio}^{\rm min}$, $f_{\gamma}$, and $f_{\rm radio}$ for any given choice of $\tau$ and $R_{\rm PRB}$.

\begin{table}[H]
 \begin{tabularx}{0.75\textwidth}{ccccc}
    \hline\hline \\[-12pt]
    $\tau$ (yr) & $F_{\gamma}^{\rm min}$ (erg/kpc$^2$/yr) & $f_{\gamma}$ & $F_{\rm radio}^{\rm min}$ (erg/kpc$^2$/yr) & $f_{\rm radio}$ \\
 
    \hline \\[-12pt]
  
    $5 \times 10^3$& $2.37 \times 10^{42}$  &0.63  & $1.33 \times 10^{41}$  & 0.27 \\

    $1 \times 10^4$& $4.22 \times 10^{42}$ &0.62 & $2.66 \times 10^{41}$ & 0.28 \\

    $2 \times 10^4$ & $9.44 \times  10^{43}$  &0.63  & $3.98 \times 10^{42}$ & 0.26 \\
    \hline\hline
  \end{tabularx}
    
\caption{\label{tab:tableS}The best-fit values of $F_{\gamma}^{\rm min}$, $F_{\rm radio}^{\rm min}$, $f_{\gamma}$, and $f_{\rm radio}$ for three values of $\tau$, and for  $R_{\rm PSR}$ = 0.01 $\, {\rm yr}^{-1}$.}
\label{t2}
\end{table}

\begin{figure}[t]
\centering \includegraphics[width=120mm]{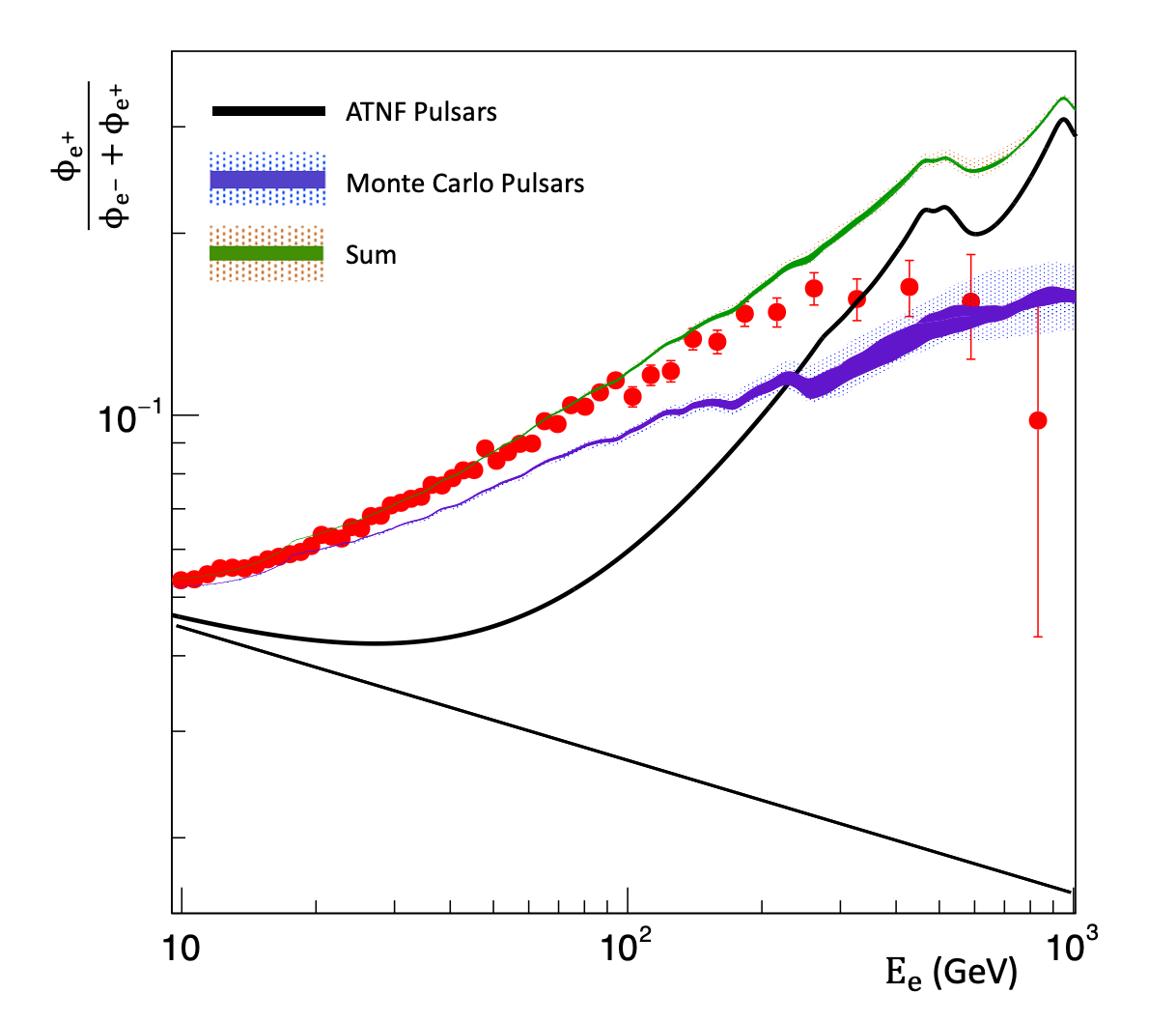} 
    \caption{The cosmic-ray positron fraction from the pulsars in the ATNF catalog (black), from unknown pulsars as calculated by our Monte Carlo (blue), and from the sum of these contributions (orange). In this figure, we have adopted the following parameters: $\alpha=1.5$, $E_c =50 \, {\rm TeV}$, $\eta=0.15$, $\tau=10^4\, {\rm yr}$, $B_{\rm ISM}=3 \, \mu{\rm G}$, and $t_{\rm PWN} = 3 \times 10^4 \, {\rm yr}$, and values of $\tau_c$, $r$, and $P$ as reported in the ATNF catalog. The solid (shaded) bands around the blue and orange curves reflect the variation observed across 68\% (all) of the realizations of our Monte Carlo.}
    \label{K9}
\end{figure}

In Fig.~\ref{K9}, we show the cosmic-ray positron fraction from the pulsars in the ATNF catalog, from unknown pulsars as calculated by our Monte Carlo, and from the sum of these contributions, adopting the best-fit values of $F_{\gamma}^{\rm min}$, $F_{\rm radio}^{\rm min}$, $f_{\gamma}$, and $f_{\rm radio}$ and otherwise using the same parameters as adopted in the previous section. In the following section, we will consideration variations of these parameters and assess their impact on our results.

\section{Parameter Variations}

\begin{figure}[t]
\centering \includegraphics[width=75mm]{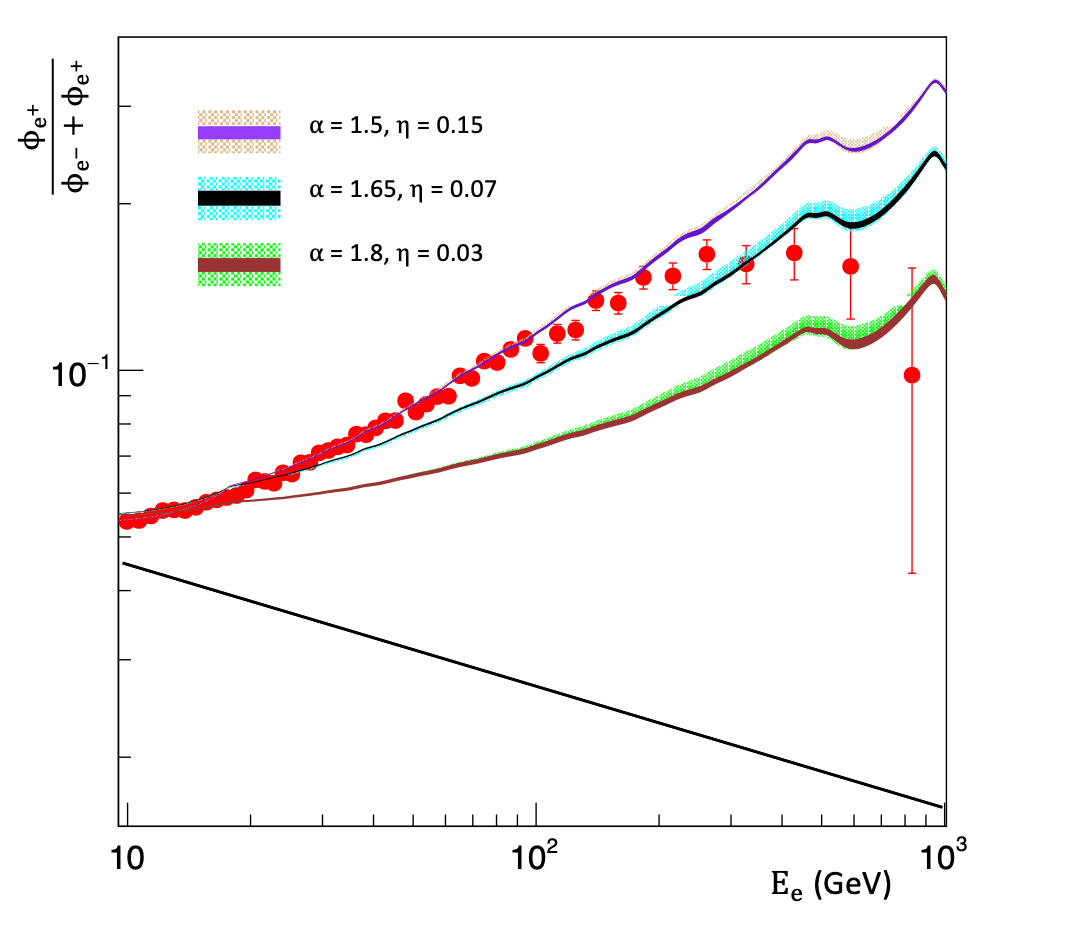}
\centering \includegraphics[width=75mm]{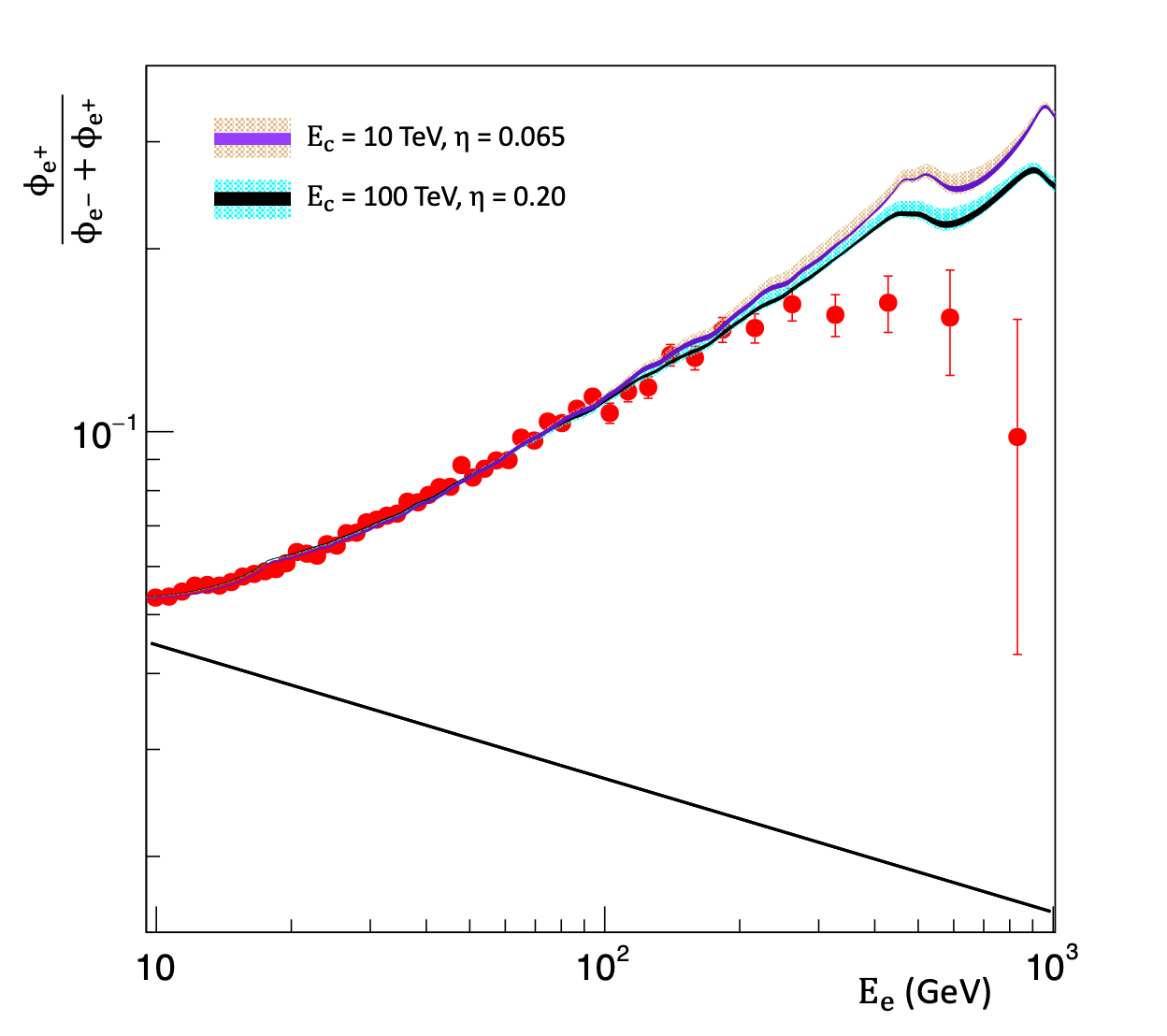}
\centering \includegraphics[width=75mm]{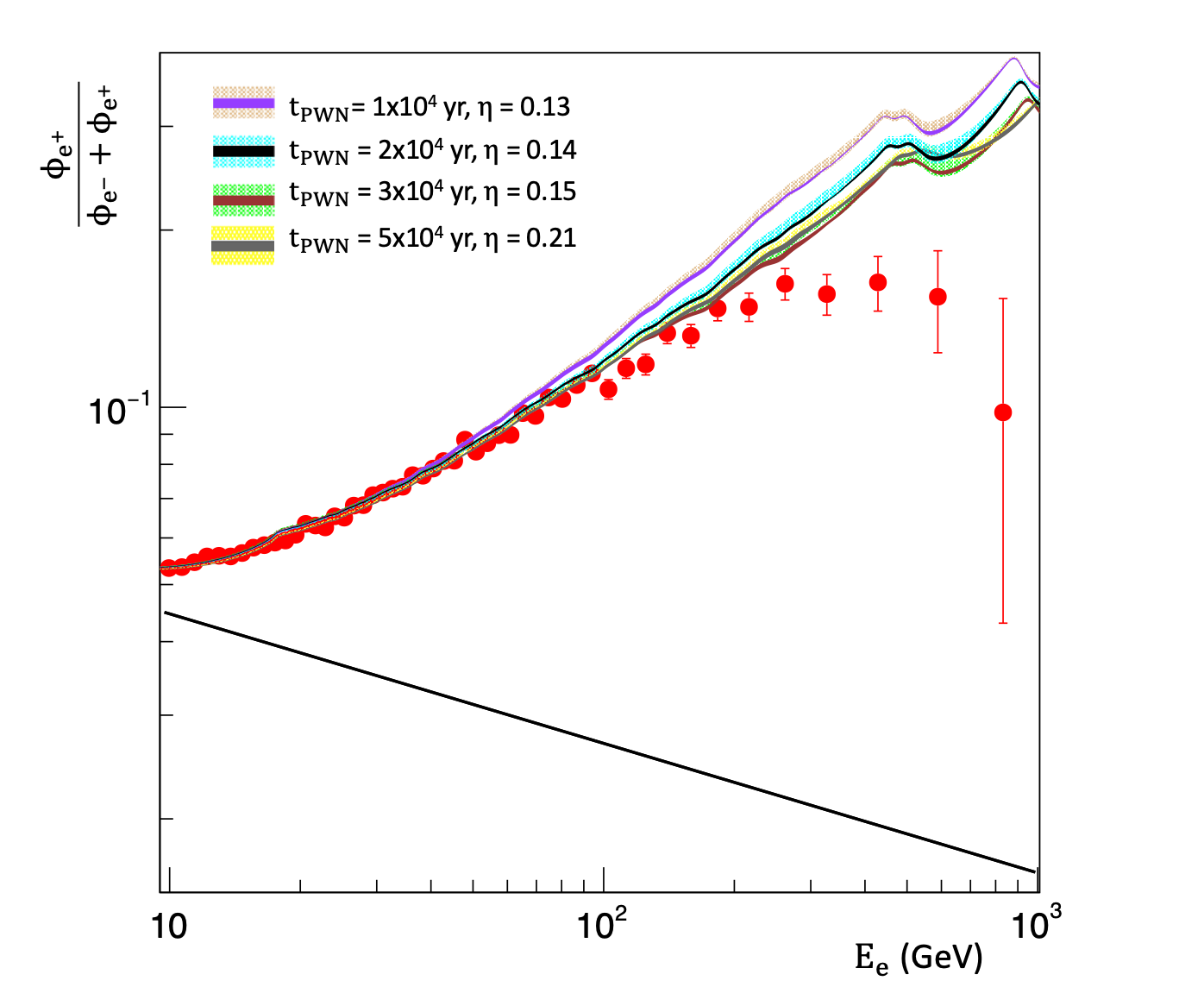}
\centering \includegraphics[width=75mm]{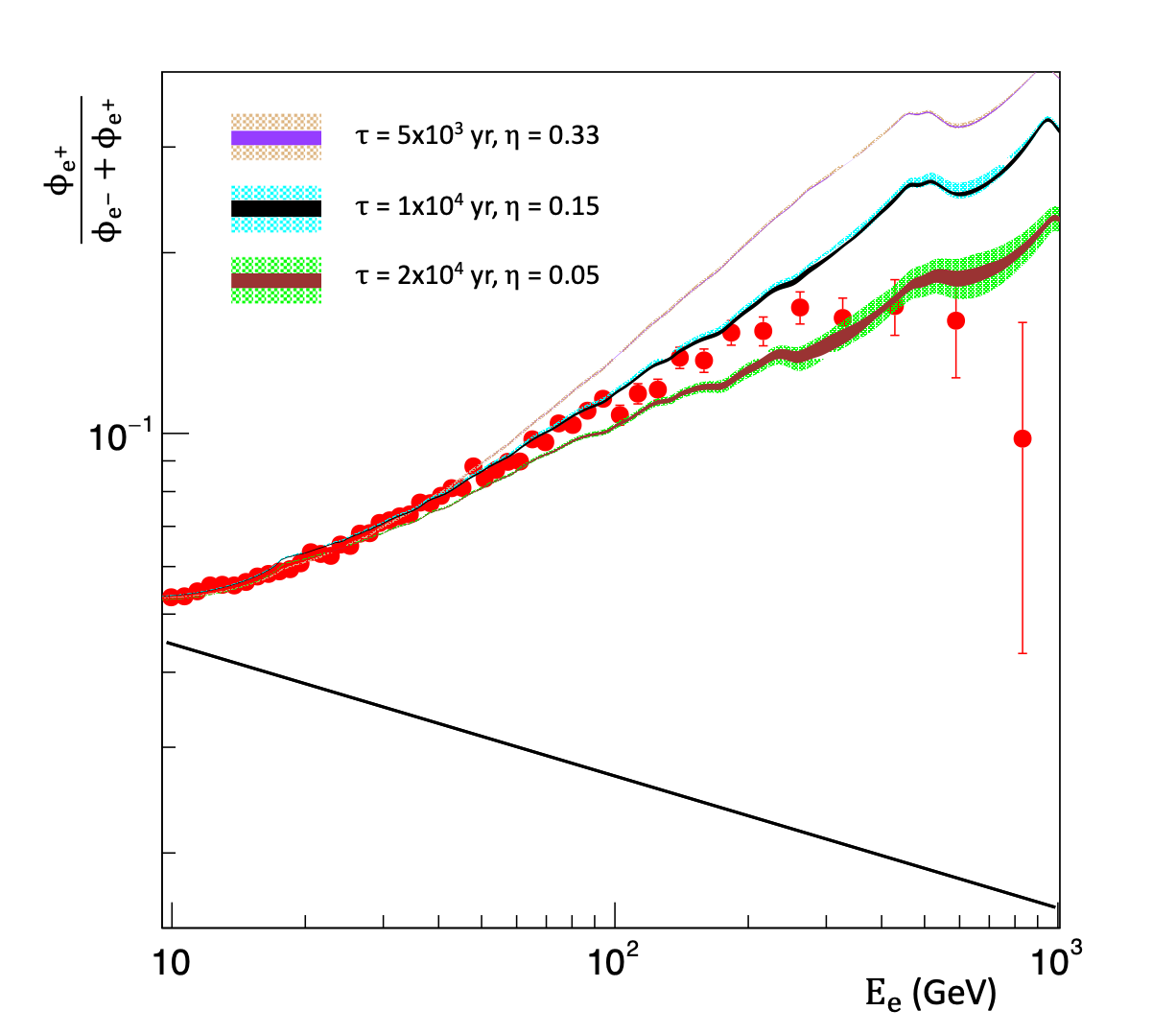}
\centering \includegraphics[width=75mm]{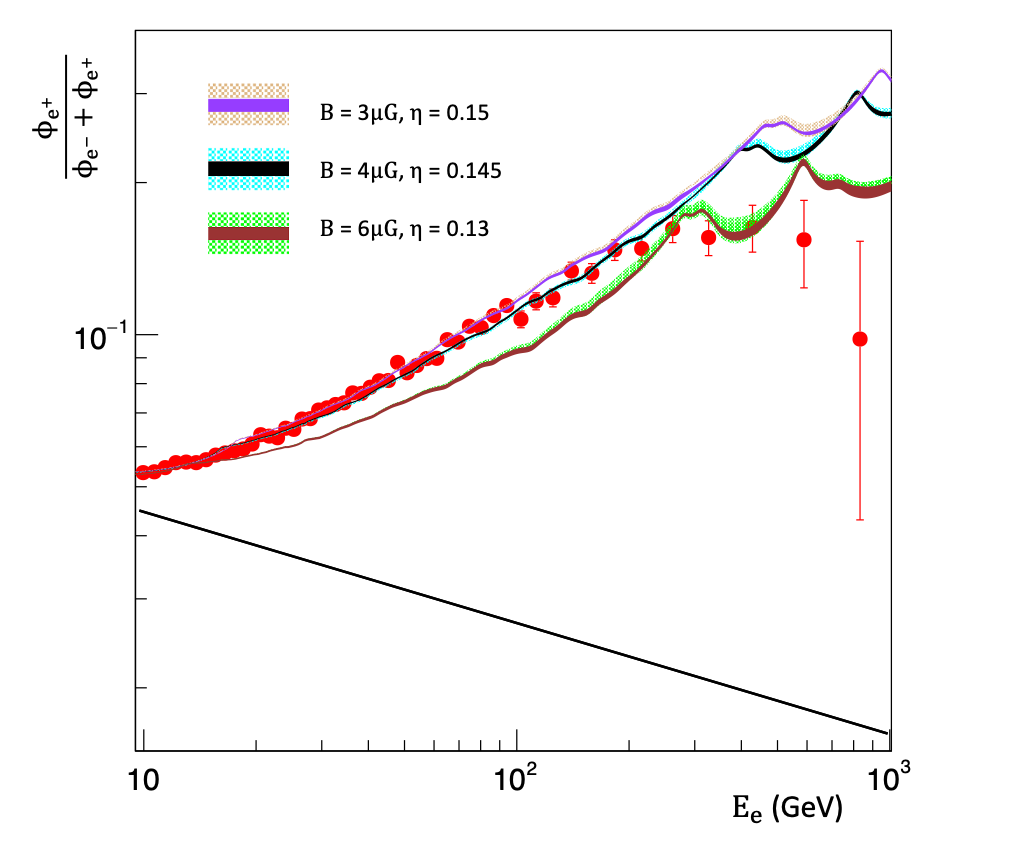}
    \caption{The impact on the positron fraction of varying the injected spectral index, $\alpha$, the injected spectral cutoff, $E_c$, the duration of the pulsars wind nebula phase, $t_{\rm PWN}$, the spindown timescale, $\tau$, or the magnetic field of the ISM, $B$.}
    \label{var1}
\end{figure}

The results shown in Fig.~\ref{K9} were obtained using our default set of parameter values: $\alpha=1.5$, $E_c=50 \, {\rm TeV}$, $t_{\rm PWN}=3 \times 10^{4}\,{\rm yr}$, $\tau=10^4 \,{\rm yr}$, $B=3 \, \mu{\rm G}$, and $\eta=0.15$. Any changes to these parameters will impact the predicted magnitude and spectral shape of the cosmic-ray positron fraction. In this section, we explore these variations and their connection to the observed features of the positron fraction. 

In Fig.~\ref{var1}, we illustrate the impact of a variety of parameter variations, modifying one of the above mentioned parameters at a time while keeping the others fixed to their default values (with the exception of $\eta$, which is selected in each case to normalize the positron fraction to its measured value at $E_e=20 \, {\rm GeV}$). Focusing first on the positron fraction as measured at energies below a few hundred GeV, the positrons in this energy range arise from many pulsars, and thus the positron fraction is largely insensitive to the location, age, or other characteristics of any specific pulsars. From this data, it is clear that an injected spectral index of approximately $\alpha \sim 1.5-1.6$ is preferred, although the precise value of this parameter also depends on the choices of $t_{\rm PWN}$, $\tau$, and $B$ that are adopted. 
For some choices of these parameters (such as $B=6\, \mu{\rm G}$, for example), the shape of the predicted positron spectrum is more concave than is measured by AMS-02, leading us to disfavor such scenarios.

In some cases, there are substantial degeneracies between the parameters we have considered. In particular, we find that the value of the injected cutoff energy, $E_c$, is approximately degenerate with the electron-positron production efficiency, $\eta$. This is demonstrated in the upper right frame of Fig.~\ref{var1}, where we show that the low-energy positron fraction is not significantly impacted by variations in $E_c$, other than in terms of the overall normalization. If we increase the value of $E_c$, we must compensate by increasing the value of the efficiency, $\eta$.

Upon reviewing these results, one might reasonably be concerned that when we select model parameters which provide a good fit to the positron fraction at low energies, we consistently predict a positron flux at high energies that is larger than that observed by AMS-02. This could be an indication that some young pulsars lose their kinetic energy more rapidly than we have assumed (in this study, we have limited our calculations to the case of magnetic dipole breaking). This could also be impacted by the stochastic nature of Inverse Compton scattering in the Klein-Nishina limit, which we have thus far treated as continuous \cite{ https://doi.org/10.48550/arxiv.2206.04699}. See our definitions in section \ref{sec:flux} and in particular equation \ref{Hooper2} for details. Furthermore, in light of the fact that the positron fraction is dominated at the highest measured energies by only a small number of sources, this could simply be the consequence of a small number of young and nearby pulsars which have characteristics that are substantially different from the average of the local population. We will explore this possibility further in the following section.

\section{Pulsar-to-Pulsar Variations}

Throughout this study, we have adopted parameter values which are intended to reflect the average characteristics of the local pulsar population. At low and moderate energies, we can expect this approach to be adequate, as the observed positron fraction is the result of non-negligible contributions from several dozen individual pulsars. At higher energies, however, the measured positron fraction could potentially be dominated by the emission from only a few individual pulsars. Judging from Fig.~\ref{g5}, for example, the single most important pulsar (PSR J1057-5226) contributes less than 15\% of the observed positrons at 100 GeV, but is responsible for producing roughly half of these particles at 500 GeV. The specific parameters for this individual pulsar are thus quite important for our prediction of the positron fraction at high energies. More generally speaking, stochastic variations in the parameters of individual pulsars make it difficult to reliably predict the positron fraction at energies above a few hundred GeV. It is, after all, very plausible that the pulsar parameters being considered in this study are not universal, but rather vary significantly from pulsar-to-pulsar. 

\begin{figure}[t]
\centering \includegraphics[width=100mm]{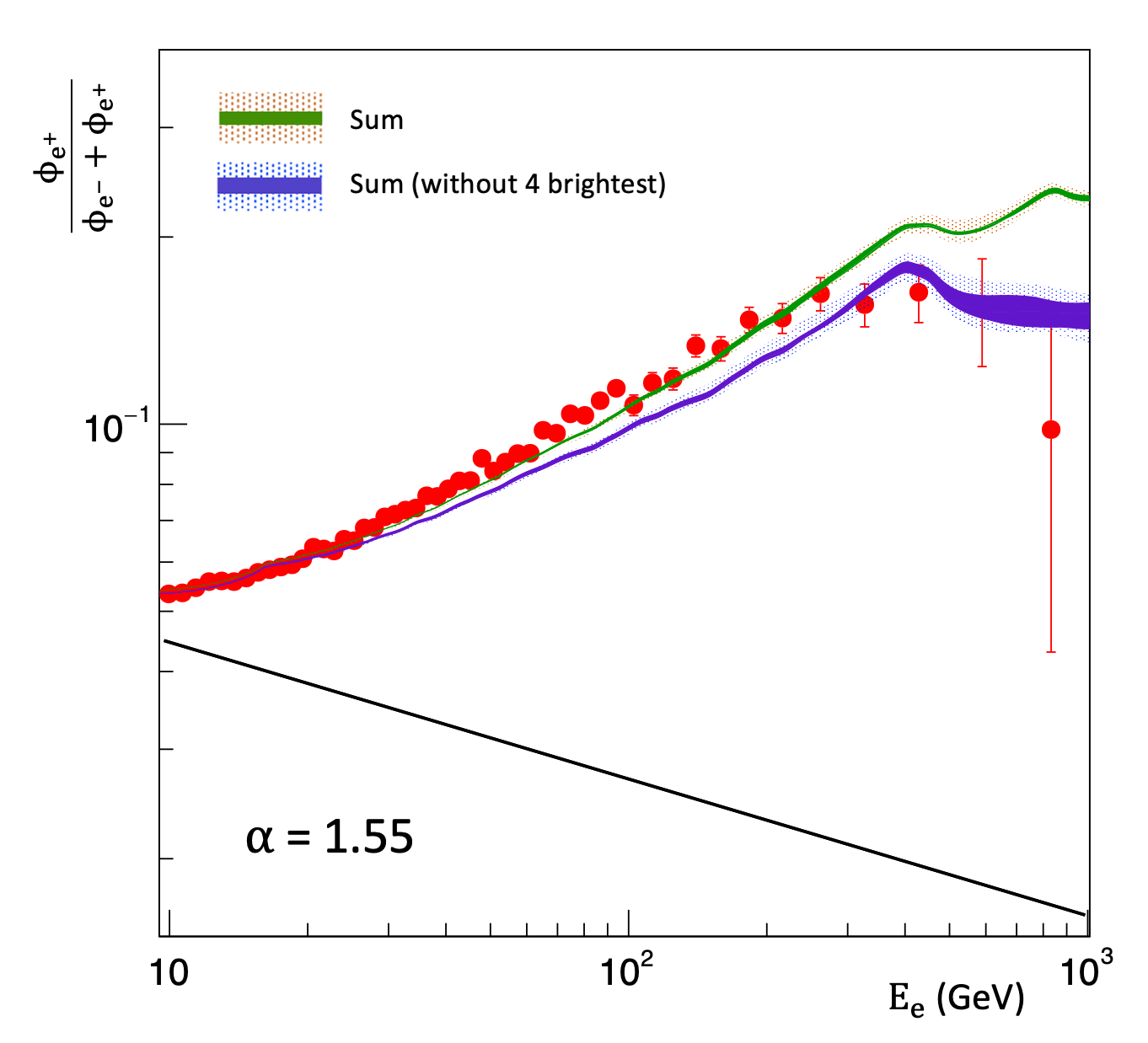} \includegraphics[width=100mm]{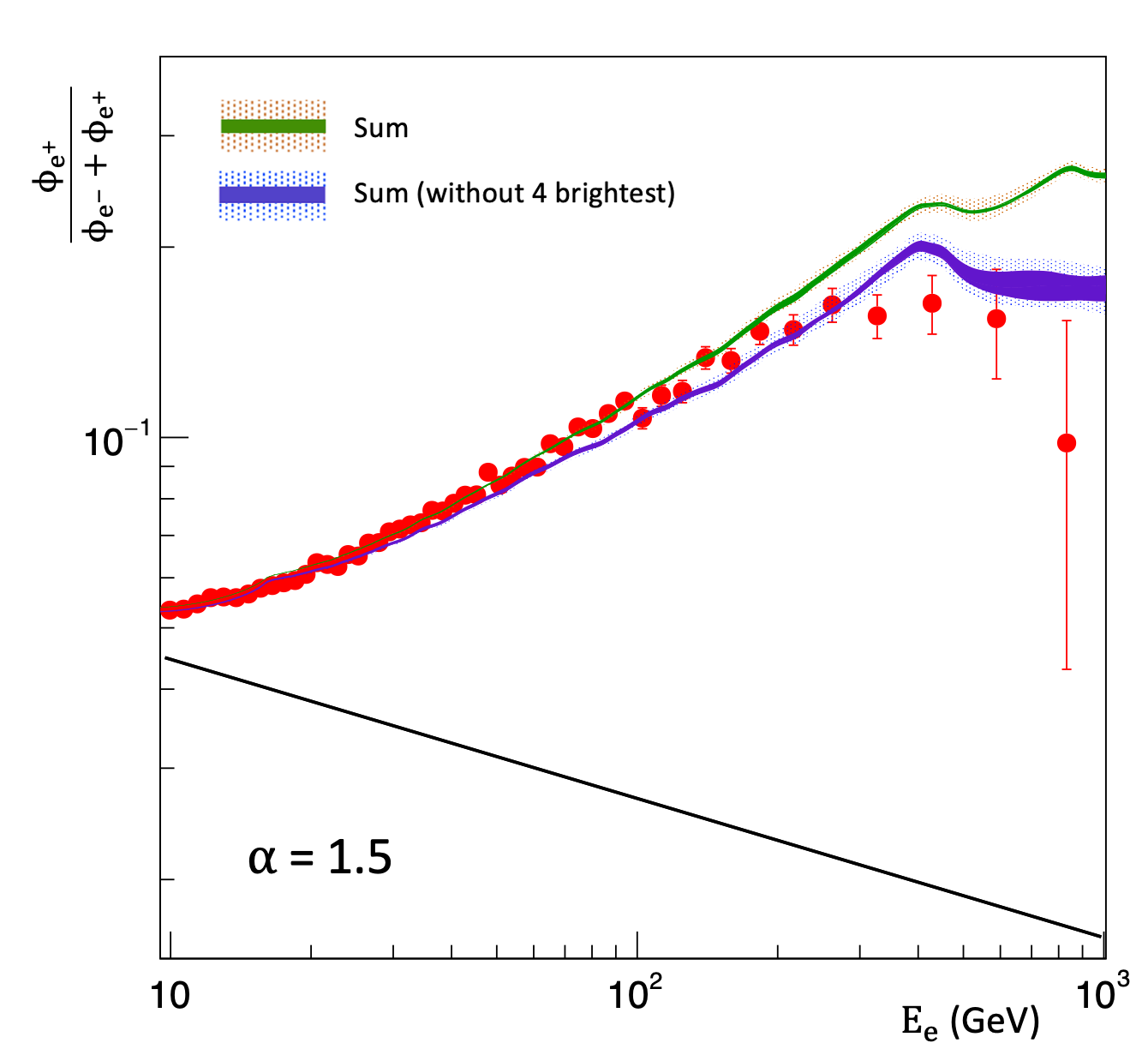}
    \caption{The cosmic-ray positron fraction as calculated for $\alpha=1.55$ (top) or $\alpha=1.50$ (bottom), $E_c=10\, {\rm TeV}$, $t_{\rm PWN} = 5 \times 10^4 \, {\rm yr}$, $\tau=10^4 \, {\rm yr}$ and $B=4 \, \mu {\rm G}$. In each frame, we show the positron spectrum that results when we include all of the pulsars (both those in the ATNF catalog, and those simulated by our Monte Carlo), as well as that obtained when we neglect the contributions from the first four pulsars listed in Table~\ref{tab:table1}. This illustrates the importance of pulsar-to-pulsar parameter variations, in particular at energies above a few hundred GeV.}
    \label{final}
\end{figure}

We explore some examples of the possible impact of pulsar-to-pulsar variations in Fig.~\ref{final} where we show the positron fraction for the case of $\alpha=1.55$ (top) or 1.50 (bottom), $E_c=10\, {\rm TeV}$, $t_{\rm PWN} = 5 \times 10^4 \, {\rm yr}$, $\tau=10^4 \, {\rm yr}$ and $B=4 \, \mu {\rm G}$. In each frame, we show the positron spectrum that results when we include all of the pulsars (both those in the ATNF catalog, and those simulated by our Monte Carlo), as well as that obtained when we neglect the contributions from the first four pulsars listed in Table~\ref{tab:table1}. At low to moderate energies, these few pulsars do not impact the positron fraction much, and we consistently obtain a result which is in reasonably good agreement with the measured spectrum. At energies above a few hundred GeV, in contrast, the result changes significantly depending on the detailed properties of these few individual pulsars. This makes it impossible for us to use our approach to reliably predict the positron fraction at the highest energies measured by AMS-02.

We acknowledge that our assumed choice of sources, especially those used to normalize to data at low energies, does lead to a unique set of best fit parameters. In fact, it is possible that one could, in principle, find a another set of "best fit" free parameters which are chosen based upon other additional sources which could reduce the aforementioned overshoot at high energies \cite{Shaviv_2009}. Also, we note that the possible impacts of spatial variations in the diffusion constant used could further play a role in how the overall contribution is modeled, particularly at high energies. However, from earlier work, it was found that lowering the diffusion coefficient too much, or indeed varying it, would negate any meaningful contributions from local pulsars \cite{Hooper:2017tkg}. 

Finally, we did explore possible further model variations, especially in comparisons of the known source contributions and the Monte Carlos. We find that while our assumptions and approximations hold up at low energies, our high energy discrepancies could be due to the fact that we do not fully model the specific pulsar to pulsar spatial variations. In particular, our Monte Carlo modeling does not take into account the variations in pulsar initial period, spin-down time, or spectral index. Future studies might benefit from further exploring and amending the pulsar modeling to account for this.

\section{Summary and Conclusions} 

Observations by the HAWC Collaboration indicate that pulsars are accompanied by TeV halos which produce a significant contribution to the cosmic-ray positron flux. In this study, we have used the measured characteristics of the cosmic-ray positron fraction to constrain the properties of the local population of pulsars and their associated TeV halos. More specifically, we have calculated the contribution to the cosmic-ray positron flux from the pulsars contained within the Australia Telescope National Facility (ATNF) pulsar catalog~\cite{Manchester_2005}, and used a Monte Carlo to estimate the contribution from those pulsars which have thus far gone undetected. After varying a number of parameters, including the spectral shape of the injected positrons, the timescale for pulsar spindown, the efficiency for electron-positron production, the magnitude of the magnetic field in the ISM, and the duration of the pulsar wind nebula phase, we have compared our results to the positron fraction as measured by AMS-02. For reasonable parameter values, we find that it is possible to obtain good agreement with the measured positron fraction, at least up to energies of $E_e \sim 300 \, {\rm GeV}$. This result further supports the conclusion that TeV halos are the main source of the observed cosmic-ray positron spectrum. At energies above a few hundred GeV, the positron fraction is dominated by a small number of pulsars, limiting our ability to reliably predict the shape of the expected positron fraction or to use this high-energy information to draw reliable conclusions pertaining to the Milky Way's broader pulsar population.

Overall, the observed positron spectrum supports a picture in which pulsars transfer $\eta \sim 5-20\%$ of their total spindown power into the production of high-energy electron-positron pairs, generating a spectrum of such particles with a hard spectral index, $\alpha \sim 1.5-1.7$. Our best fits were obtained for models in which the radio beams of pulsars cover $\sim 28\%$ of the surrounding solid angle, while the gamma-ray beams are significantly wider, covering $\sim 62\%$ of the total solid angle.

\section*{Acknowledgments} 

We would like to thank Tim Linden for helpful discussions. DH is supported by the Fermi Research Alliance, LLC under Contract No.~DE-AC02-07CH11359 with the U.S. Department of Energy, Office of Science, Office of High Energy Physics. 


\printbibliography[heading=bibintoc]

\end{document}